# Electric Field Switching of Magnon Spin Current in a Compensated Ferrimagnet

Kaili Li, Lei Wang, Yu Wang*, Yuanjun Guo, Shuping Lv, Yuewei He, Weiwei Lin*, Tai Min,

Shaojie Hu, Sen Yang*, Dezhen Xue, Aqun Zheng, Shuming Yang*, Xiangdong Ding

K. Li, Y. Wang, Y. Guo, S. Lv, Y. He, T. Min, S. Hu, S. Yang, D. Xue, X. Ding
MOE Key Laboratory for Nonequilibrium Synthesis and Modulation of Condensed Matter and
State Key Laboratory for Mechanical Behavior of Materials
School of Physics
Xi'an Jiaotong University
Xi'an 710049, China
e-mail: yuwang@mail.xjtu.edu.cn; yangsen@mail.xjtu.edu.cn

L. Wang, W. Lin
Key Laboratory of Quantum Materials and Devices of Ministry of Education
School of Physics
Southeast University
Nanjing 211189, China
wlin@seu.edu.cn

A. Zheng
*School of Chemistry,*
*Xi'an Jiaotong University*
*Xi'an 710049, China*

S. M. Yang
*State Key Laboratory for Manufacturing Systems Engineering,*
Xi'an Jiaotong University
Xi'an, 710049, China
shuming.yang@mail.xjtu.edu.cn

**Manipulation of directional magnon propagation, known as magnon spin current, is essential for developing magnonic memory and logic devices featuring nonvolatile functionalities and ultralow power consumption. Magnon spin current can usually be modulated by magnetic field or current-induced spin torques. However, these approaches may lead to energy dissipation caused by Joule heating. Electric-field switching of magnon spin current without charge current is highly desired but very challenging to realize. By integrating magnonic and piezoelectric materials, we demonstrate manipulation of the magnon spin current generated by the spin Seebeck effect in the ferrimagnetic insulator $Gd_3Fe_5O_{12}$ (GdIG) film on a piezoelectric substrate. We observe reversible electric-field switching of magnon polarization without applied charge current. Through strain-mediated magnetoelectric coupling, the electric field induces the magnetic compensation transition between two magnetic states of the GdIG, resulting in its magnetization reversal and the simultaneous switching of magnon spin current. Our work establishes a prototype material platform that pave the way for developing magnon logic devices characterized by all electric field reading and writing and reveals the underlying physics principles of their functions.**

1. Introduction

The magnon spin current or spin wave is the intrinsic excitation of spin dynamics in a magnetically ordered media.[1,2] In contrast to an electric current, magnon spin current can propagate in an insulator without the flow of charge carriers; thus, Joule heating—the major cause of energy dissipation in modern electronics—is avoided, bringing out a thriving new research field of magnonics that offers a promising route towards ultralow-power information processing.[3-11]

However, development of magnonic device as a viable solution for future logic operation still faces many challenges. A fundamental problem is the lack of demonstration of considerable signal switching of magnon spin current by an energy-efficient approach for non-volatile information storage and processing.[12-18] The manipulation of magnon spin current can typically be achieved by applying a magnetic field,[19,20] electric current,[21-24] and electric field,[25-29] among which electric field manipulation is considered the best choice owing to the advantage of ultralow energy consumption. [30-32] However, the electric field effect on magnon spin current is usually of small modulation amplitude via magnetic anisotropy,[25,26] antiferromagnetism,[27,28] interfacial magnetism,[29] and Dzyaloshinskii-Moriya-like spin-orbit interaction,[33] or exists at extremely low temperature via Rashba-Edelstein effect,[30] which is easy to cause signal mixing and far from the resolution and temperature requirements of magnonic devices. A convincing demonstration of electric field-induced full switching of magnon spin current at ambient temperature is still lacking. Here, we achieved an unambiguous reversible switching of magnon spin current by applying an electric field in the compensated ferrimagnetic insulator at ambient temperature and small magnetic fields.

The compensated ferrimagnetic insulators REIG (RE = Gd, Tb, Dy, Ho, Er) are a class of material with the garnet crystal structure (see Supplementary Information).[34] This structure consists of three sublattice magnetizations: tetrahedrally coordinated Fe atoms ($d$-sites), octahedrally coordinated Fe atoms ($a$-sites), and dodecahedrally coordinated RE atoms ($c$-sites). As depicted in the schematic phase diagram of **Figure 1**a, a magnetic compensation transition occurs at $T_{\text{comp}}$, and the sublattice magnetization reverses between the two magnetic states below and above $T_{\text{comp}}$.[34] In particular, magnon spin current flips has been observed at this magnetic compensation transition.[35,36] On the other hand, external stress or strain can also induce such a magnetic compensation transition because this transition accompanies a sudden change in lattice constants and thus couples with the external mechanical field.[37,38] Notably, the lattice change of compensated ferrimagnetic insulator is close to the electrostrain of piezoelectrics.[34,39] This feature paves a promising way for the electric field switching of magnon spin current in such materials.

## 2. Results and Discussions

To verify the feasibility of electric field switching of magnon spin current, we designed a Pt (5 nm)/GdIG(72 nm)/MgO(6.5 nm)/PMN-PT heterostructure as illustrated in **Figure 1**b. The (011)-oriented PMN-PT piezoelectric single crystal is used as the substrate to generate considerable electro-strain along the [100] direction, with an electrical gate field $E_G$ applied between the Pt layer and the bottom electrode. We selected GdIG as the compensated ferrimagnetic insulator because it exhibits the highest $T_{comp}$ among the REIG materials. [34] The MgO buffer layer is inserted between the GdIG and the PMN-PT to reduce the lattice mismatch and prevent the Pb in the PMN-PT substrate to diffuse into the GdIG film. The magnon spin current generated by the spin Seebeck effect (SSE) is dominated by the Fe sublattice magnetization rather than the net magnetization of their three sublattices in GdIG. [36] Due to this unique property, the temperature-induced reversal of magnon spin current is only observable by SSE, not by spin pumping or spin Hall magnetoresistance. [40,41] Thus, in the present work, the thermal magnon spin current under a temperature gradient emerges through SSE and is detected in heavy matter Pt via inverse spin Hall effect (ISHE). [42] As illustrated in **Figure 1**b, applying $E_G$ along $z$ or [011] direction to the heterostructure is equivalent to exerting an electrostrain (compression strain along $x$ or [100] direction and tension strain along $y$ or [01$\bar{1}$] direction) of PMN-PT to the GdIG thin film. Such an electrostrain may induce a transition between the low- and high-temperature magnetic states (**Figure 1**a), at which the sublattice magnetization reverses, accompanied by switching of magnon spin current.

The high-resolution TEM image for the cross-section of the Pt/GdIG/MgO/PMN-PT heterostructure is shown in **Figure 2**a. It reveals that the GdIG film is highly crystalline, with a smooth interface to the MgO, facilitating electrostrain transmission efficiency between the GdIG and the PMN-PT substrate. Based on the magnetization vs. temperature ($M$-$T$) curve (see Supplementary Information), the $T_{comp}$ of the studied sample is determined to be 279 K, which is slightly lower than that observed in bulk samples due to Fe deficiency and lattice mismactch. [38] The $V_{ISHE}$ vs. $H$ curves of the sample were measured at temperatures extending from above $T_{comp}$ to below $T_{comp}$; the selected curves obtained at $T_{Pt}$ = 315, 279, 255 K are shown in **Figure 2**b. The SSE voltage ($V_{SSE}$), defined as $V_{SSE}$ = $\frac{1}{2}[V_{ISHE}(+\mu_0 H_{sat}) - V_{ISHE}(-\mu_0 H_{sat})]$, can be obtained from these $V_{ISHE}$ vs. $H$ curves, where the $\mu_0 H_{sat}$ is the saturation magnetic field of the GdIG. To clearly reveal the temperature evolution of SSE, we plot the SSE voltage ($V_{SSE}$) as a function of temperature in **Figure 2**c. As shown in **Figure 2**c, the sign of $V_{SSE}$ is negative above $T_{comp}$ (= 279 K), with its magnitude gradually decreasing upon cooling, approaching zero around $T_{comp}$. On further cooling below $T_{comp}$, $V_{SSE}$ becomes positive and its magnitude increases with decreasing temperature. The net magnetization of the GdIG is dominated by the $d$-site Fe magnetization at $T > T_{comp}$, which is along the applied magnetic field direction but opposite to the magnetization direction of the Gd and the $a$-site Fe (see Supplementary Information). The magnetization of the Gd sublattice increases dramatically upon cooling, and the total magnetization of the Gd and the *a*-site Fe becomes dominated as the temperature decreases to $T < T_{comp}$. Their magnetization reverses orientation and aligns along the magnetic field direction to reduce the Zeeman energy. Simultaneously, the *d*-site Fe sublattices magnetization flips its direction as shown in **Figure 1**a and Supplementary Information. The magnon current general by SSE in GdIG are dominated by the magnetization of *d*-site Fe sublattice near the magnetic compensation transition. The reversal of *d*-site Fe magnetization during the magnetic compensation transition leads to the sign change of $V_{SSE}$ and magnon spin current at $T_{comp}$, which establishes a well-suited platform for manipulating magnon spin current.

We next turn to the effect of electric field (*E*) on the SSE of the Pt/GdIG/MgO/PMN-PT sample. **Figure 3**a shows the $V_{ISHE}$-*H* loops measured with various $E_G$ at $T_{Pt}$ = 274 K, slightly below $T_{comp}$. The amplitude and squareness ($V_{ISHE(H = 0)}/V_{ISHE(H = 0.6\ T)}$) of $V_{ISHE}$-*H* loops decrease when *E* reduces from +1 kV/cm to −0.5 kV/cm, and the $V_{ISHE}$-*H* loop inverses when *E* further decreases to −1 kV/cm. Moreover, we present the $V_{SSE}$ as a function of *E* at $T_{Pt}$ = 274 K, as shown in **Figure 3**b. The sign of $V_{SSE}$ reverses as the *E* changes from the electrical field of +1 kV/cm to −1 kV/cm, namely, the electric field fully inverses the sign of magnon spin current. This finding indicates that the electric field can induce the reorientation of *d*-side Fe sublattice magnetization and reverse the magnon spin current polarization. **Figure 3**b reveals that the sign of two distinctive magnon spin current states are not coincident at *E* = 0 kV/cm, demonstrating that the electrical field tunable signal is non-volatile and reversible. As shown in **Figure 3**c, by applying *E* = −1 kV and *E* = +1 kV/cm at a magnetic field of −0.5 T, $V_{SSE}$ toggles between negative $V_{SSE}$ ≈ +0.5 μV and positive $V_{SSE}$ ≈ −0.9 μV. Compared with amplitude modulation, the apparent signal switching avoids information confusion and can be easily identified, which is beneficial for manufacturing electric-field-controlled magnonic devices.

To understand the electric field-induced magnon spin current switching, we carry out the first principle calculations on bulk GdIG as shown in **Figure 4**a-c. As mentioned above, the unit cell of GdIG comprises four nonequivalent atoms, marked as $a$-Fe$^O$, $d$-Fe$^T$, $c$-Gd, O in **Figure 4**a. Given that the magnetization of O atoms is negligible, considerable exchange interactions occur mainly between $a$-Fe$^O$, $d$-Fe$^T$ and $c$-Gd atoms, and thus we have the exchange interactions $J_{ij}$, $i, j \epsilon \{a,d,c\}$. Following the parameter space of Ref. [39], our simulation yields $J_{aa}$ = 0.083 meV, $J_{dd}$ = 0.132 meV, $J_{ad}$ = 2.451 meV, $J_{ac}$ = −0.024 meV, and $J_{dc}$ = 0.189 meV, consistent with the values reported in Ref. [43]. Under an electric field, an in-plane electrostrain will be applied on GdIG from the PMN-PT substrate. The lattice constant is stretched by an amount of $\varepsilon_x$ = −2116 ppm along the $x$-axis and $\varepsilon_y$ = +813 ppm along the $y$-direction, according to the experimental data of the in-plane electrostrain for PMN-PT.[44] With this condition, the optimal thickness is obtained by calculating the total energy of the GdIG with varying lattice constant in the $z$-direction, which turns out to be a $z$-axis deformation of $\varepsilon_z$ = 1.3%, as shown in **Figure 4**a. We then calculate the $J_{ij}$ under these conditions and find all $J_{ij}$ values are similar to the results for the bulk GdIG. Therefore, we conclude that the exchange interactions in GdIG are not sensitive to the small strain resulting from the substrate. However, the original crystal symmetry ($Ia\bar{3}d$) of GdIG is broken due to electrostrain, leading to a change in the spin-wave spectrum. As shown in **Figure 4**b, c, the energy of the spin-wave spectrum composed of $\alpha$, $\beta$ and $\gamma$ modes is depressed by electrostrain. As reported by previous studies, the SSE signal is dominated by the $\beta$ mode around the magnetic compensation transition, and the depression of the $β$ mode (**Figure 4**c) stabilizes the high-temperature magnetic state of GdIG. [36,41]

Thus, the results of **Figure 4**b, c suggest that the applied electric field (or electrostrain) can change the low-temperature magnetic phase into the high-temperature magnetic phase in GdIG and shift the corresponding $T_{comp}$ to lower temperature.

To further verify this point, we perform Monte Carlo simulations similar to that reported in Ref. [43,45] to calculate the temperature dependence of the magnetizations in GdIG. **Figure 4**d shows the calculated magnetization vs. temperature (*M-T*) curve around the $T_{comp}$, which reveals that the $T_{comp}$ shifts to a lower temperature by 3 K under the electrostrain. This simulation result is further confirmed by the $V_{SSE}$ vs. *T* curves obtained at different $E_G$ (**Figure 4**e), which are obtained by measuring the dependence of $V_{SSE}$ with $E_G$ at different temperatures slightly above and below $T_{comp}$. The $T_{comp}$ (determined by the $V_{SSE}$ vs. *T* curve at $V_{SSE} = 0$) shifts to a lower temperature by more than 9 K when changing $E_G$ from 0 to −1 kV/cm. Both simulation and experimental results demonstrate that the transition between high-and low-temperature magnetic states can take place by applying electrostrain within a considerable temperature window around $T_{comp}$. The magnon current polarization (i.e., *d*-side Fe sublattice magnetization) of GdIG reverses between the high- and low-temperature states, allowing the applied electric field to switch the associated magnon spin current.

## 3. Conclusion

In summary, we have demonstrated that the magnon spin current can be fully switched by an applied electric field in the compensated ferrimagnet-based Pt/GdIG/MgO/PMN-PT heterostructure. Such a novel effect is originated from the electric field induced magnetic compensation transition between two magnetic states with opposite magnetic polarization in GdIG, enabled by the strain-mediated magnetoelectric coupling between the GdIG film and PMN-PT substrate. Such an electric-field-controlled magnon spin current reversal is characterized by the SSE voltage signal flipping, which enables all electric reading and writing without applied charge current for information storage and processing. The underlying principle established by this work is generic and can be extended to other compensated ferrimagnet materials, opening an innovative route for energy-efficient and non-volatile magnonic devices.

## 4. Experimental Section

Sample fabrication and characterization: Hight-quality, crystalline Pt (5 nm)/ GdIG(72 nm)/ MgO(6.5 nm) trilayers were deposited on (011)-oriented PMN-PT (0.2 mm) by magnetron sputtering with a base vacuum pressure of $1.9\times10^{-7}$ mbar. The MgO buffer layer was deposited by RF-magnetron sputtering from a MgO target under the ambient of argon and oxygen mixing gas (the ratio 13:1) with the fixed pressure of $8\times10^{-3}$ mbar at a substrate temperature of 473 K. A 72 nm thick GdIG film was RF sputtered on the MgO layer from a GdIG target at a substrate temperature of 773 K, the argon and oxygen mixing gas (the ratio 3:1) is $1\times10^{-2}$ mbar. After deposition, the sample was annealed in air at 1123 K for 0.5 hour to increase the crystallinity of GdIG. Finally, a $5\times2$ mm$^2$ Pt strip was deposited on the GdIG film by DC-magnetron sputtering with a shadow mask at room temperature. The deposition rates were determined to be 0.041 nm/s, 0.039 nm/s and 0.004 nm/s for Pt, GdIG and MgO, respectively.

Transmission electron microscopy: The cross-sectional high-resolution transmission electron microscopy (TEM) observation were performed using a transmission electron microscope (Spectra 300).

Magnetic measurements: The magnetic properties were measured by a Quantum Design MPMS (MPMS-SQUID VSM-09), and magneto-thermoelecteic transport measurements were performed in a cryostat with a superconducting magnet. The gate voltage was applied by the Keithley 2400 sourcemeter. The inverse spin Hall voltage was measured by the Keithley 2182A nanovoltmeter.

Spin Seebeck measurement: The SSE measurement configuration for the Pt/GdIG/MgO/PMN-PT sample is depicted in **Figure 1**b. [42] The sample was sandwiched between a copper holder and a resistive chip heater to induce a temperature gradient ($\nabla T$) along the $z$ (out-of-plane) direction, which generates the magnon spin current in GdIG. The temperature of the copper holder ($T_{\text{copper}}$) was measured by a thermocouple, and the temperature of the Pt layer ($T_{\text{Pt}}$) was measured via its electrical resistivity. The $\nabla T$ across the sample is estimated to be 10 K/mm by the value of $T_{\text{Pt}} - T_{\text{copper}}$ and the thickness of PMN-PT substrate. The magnetic field $H$ was applied along $x$ direction. The injected magnon spin current is converted to a voltage in Pt via ISHE following $V_{\text{ISHE}} \propto \theta_{\text{SH}} \rho J_s \times \sigma$ for signal detection. The separation of the voltage contacts is about 5 mm. [42]

Calculation: The total energy of the magnetic insulator GdIG is given by the Heisenberg model written as

$$E_{tot} = E_0 - \frac{1}{2}\sum_{i \neq j} J_{ij} \boldsymbol{S}_i \cdot \boldsymbol{S}_j$$

where $E_0$ is the energy excluding spin-spin interactions, $J_{ij}$ is the exchange interactions between the atomic sites *i* and *j* and $S_{i/j}$ represents the corresponding spin vectors. By calculating the total energy of GdIG in different magnetic configurations, the exchange interactions and the spin-wave spectrum can be obtained. [45] The total energies are calculated by the Vienna ab initio simulation package (VASP), [46,47] in which the generalized gradient approximation with on-site Coulomb correction (GGA+U) and projector augmented wave (PAW) pseudopotentials are used. [48] The parameters of U-J are set to 3.3 eV and 4.7 eV for Gd and Fe atoms, respectively. A 500 eV plane-wave cutoff energy and a 6×6×6 Monkhorst-Pack *k*-point mesh are used in the calculation to optimize the trade-off between calculation speed and accuracy.

**Supporting Information**

Supporting Information is available from the Wiley Online Library or from the author.

**Acknowledgements**

K. Li, L. Wang contributed equally to this work. Y.W. acknowledge financial support from the National Natural Science Foundation of China (Grant No. 51931004, 51471127), National Key R&D Program of China (2021YFB3802102), Key research and development program of Shaanxi province (2023-ZDLGY-21). W.L. acknowledge financial support from the National Natural Science Foundation of China (Grant No. 12074065). S.Y. acknowledges financial support from the National Key R&D Program of China (2021YFB3501401), Key Scientific and Technological Innovation Team of Shaanxi Province (2020TD-001). S.M.Y. acknowledges financial support from the China National Funds for Distinguished Young Scientists (No. 52225507).

**Competing interests**

The authors declare no competing interests.

**Data availability**

The data that support the findings of this study are available from the corresponding author upon reasonable request.

**Figure 1**

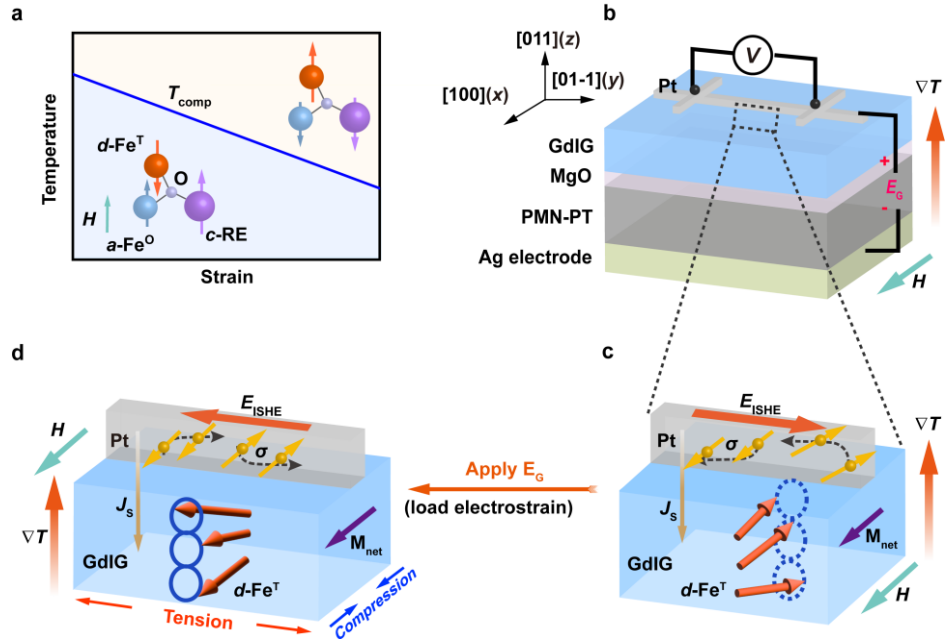

**Figure 1 | Concept of electric field switching of magnon spin current. a,** Schematic temperature-strain phase diagram of a compensated ferrimagnetic insulator, which shows that both temperature and strain can induce a transition between the two different magnetic states with opposite magnetic polarization: *i*) high-temperature magnetic state at $T > T_{comp}$, in which the net magnetization is dominated by the *d*-site Fe magnetization aligning along the applied magnetic field direction but opposite to the magnetization of the Gd and *a*-site Fe; *ii*) low-temperature magnetic state at $T < T_{comp}$, in which the net magnetization of the Gd and the *a*-site Fe becomes dominated and reverses to align parallel to the magnetic field direction. **b,** Sample geometry of the Pt/GdIG/MgO/PMN-PT system for detecting switchable magnon spin current under a gate electric field ($E_G$). **c and d,** Mechanism for the magnon spin current switching manipulated by $E_G$. When an electric field applying along the *z* or [011] direction, the PMN-PT substrate produces a compression strain along the *x* or [100] direction and tension strain along the *y* or [01$\bar{1}$] direction, which may induce the reversion of sublattice magnetizations in GdIG according to the phase diagram of **Figure 1**a, and thus leads to the sign flip of magnon spin current.

**Figure 2**

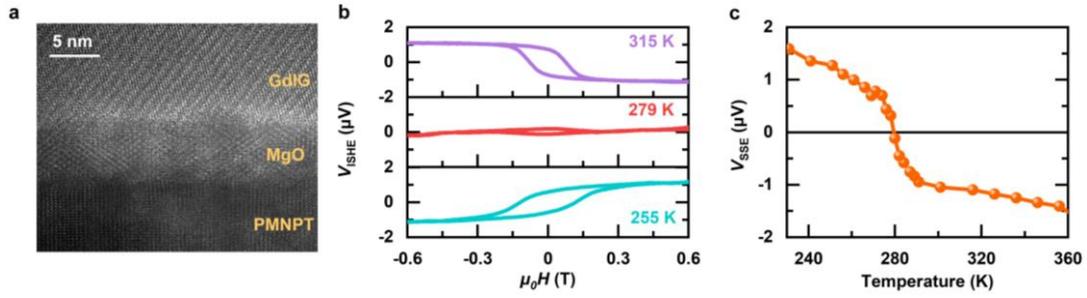

**Figure 2 | Structural characterization and SSE measurements of the sample. a,** The cross-sectional high-resolution transmission electron microscopy (TEM) image of the GdIG(72 nm)/MgO(6.5 nm)/PMNPT interfaces. **b,** $V_{ISHE}$ vs. magnetic field ($V_{ISHE}$-$H$) curves of the poled Pt/GdIG/MgO/PMN-PT sample measured from 315 K to 255 K. **c,** Temperature dependence of the SSE signal $V_{SSE} = \frac{1}{2}[V_{ISHE}(+\mu_0 H_{sat}) - V_{ISHE}(-\mu_0 H_{sat})]$ at the saturation magnetic field of 0.6 T, which shows a sign flip as the temperature is swept through the $T_{comp}$ at 279 K.

**Figure 3**

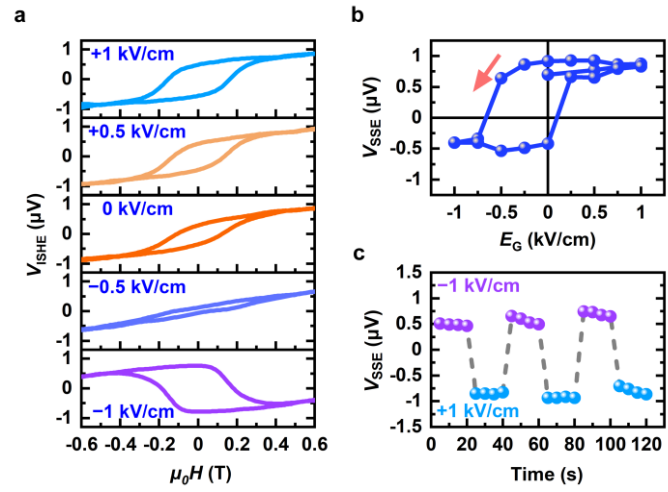

**Figure 3 | Effect of electric field (*E*) on the SSE a,** $V_{ISHE}$ vs. *H* curves of the poled sample measured under $E_G$ of +1, +0.5, 0, −0.5, −1 kV/cm at 274 K. The $V_{ISHE}$-*H* loop reverses as the $E_G$ decreases from +1 kV/cm to −1 kV/cm. **b,** SSE signal $V_{SSE}$ as a function of $E_G$ at 274 K and 0.6 T magnetic field, which denotes the electric-field-induced $V_{SSE}$ signal change from positive to negative. **c,** Reversible $E_G$-impulse-induced $V_{SSE}$ switching with remarkable amplitude under a magnetic field of −0.5 T.

**Figure 4**

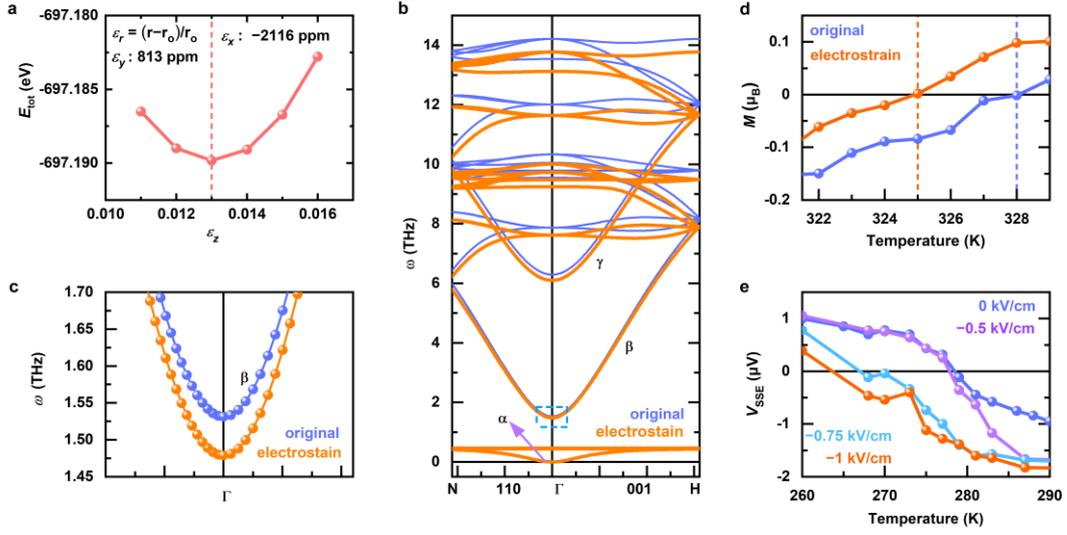

**Figure 4 | Calculated energy, spin wave spectrum and magnetization of GdIG and experimental $V_{SSE}$ vs. temperature curves. a,** Total energy of the GdIG vs. strain in the $z$ direction ($\varepsilon_z$) with a finite in-plane electrostrain ($\varepsilon_x = -2116$ ppm; $\varepsilon_y = +813$ ppm). **b,** Spin wave spectrum of the GdIG with or without strain, consisting of the $\alpha$, $\beta$ and $\gamma$ modes. **c,** Amplified plot of the area outlined by the dashed box in **Figure 4**b, which shows that the energy of the spin-wave spectrum is depressed by electrostrain. **d,** $M$-$T$ curves with and without electrostrain obtained by Monte Carlo simulation for GdIG. With the electrostrain of $\varepsilon_z = 1.3\%$, $T_{comp}$ is lowered by 3 K. **e,** $V_{SSE}$ as a function of temperature near $T_{comp}$ for various $E_G$, which experimentally confirms that the applied $E_G$ shifts $T_{comp}$ to a lower temperature.

# Supplementary Information

# Electric Field Switching of Magnon Spin Current in a Compensated Ferrimagnet

Kaili Li, Lei Wang, Yu Wang*, Yuanjun Guo, Shuping Lv, Yuewei He, Weiwei Lin*, Tai Min,

Shaojie Hu, Sen Yang*, Dezhen Xue, Aqun Zheng, Shuming Yang*, Xiangdong Ding

K. Li, Y. Wang, Y. Guo, S. Lv, Y. He, T. Min, S. Hu, S. Yang, D. Xue, X. Ding
MOE Key Laboratory for Nonequilibrium Synthesis and Modulation of Condensed Matter and State Key Laboratory for Mechanical Behavior of Materials
School of Physics
Xi'an Jiaotong University
Xi'an 710049, China
e-mail: yuwang@mail.xjtu.edu.cn; yangsen@mail.xjtu.edu.cn

L. Wang, W. Lin
Key Laboratory of Quantum Materials and Devices of Ministry of Education
School of Physics
Southeast University
Nanjing 211189, China
wlin@seu.edu.cn

A. Zheng
*School of Chemistry,*
*Xi'an Jiaotong University*
*Xi'an 710049, China*

S. M. Yang
*State Key Laboratory for Manufacturing Systems Engineering,*
Xi'an Jiaotong University
Xi'an, 710049, China
shuming.yang@mail.xjtu.edu.cn

The compensated ferrimagnetic insulators REIG (RE = Gd, Tb, Dy, Ho, Er) comprise three magnetic sublattices, as depicted in **Figure S1**a. The tetrahedrally coordinated Fe atoms ($d$-sites) and octahedrally coordinated Fe atoms ($a$-sites) are strongly coupled via antiferromagnetic superexchange interaction, while dodecahedrally coordinated RE atoms ($c$-sites) are coupled to the $a$-site Fe atoms through weak ferromagnetic exchange interaction.[1,2]

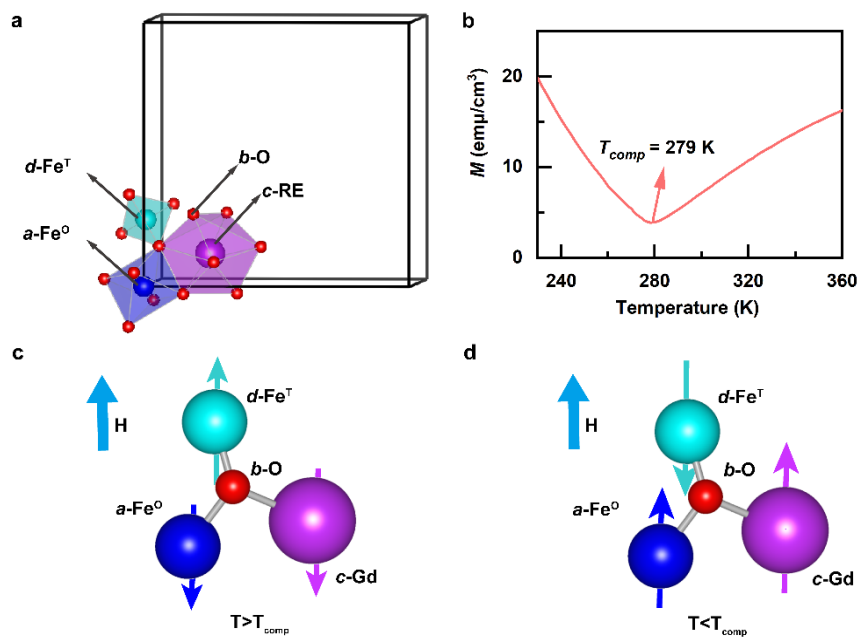

**Figure S1 | Crystal structure of REIG, temperature dependence of magnetization and magnetic sublattices of GdIG. a,** Garnet crystal structure of the REIG. **b,** Magnetization vs. temperature ($M$-$T$) curve of GdIG measured at a magnetic field of 0.3 T. The magnetic compensation temperature ($T_{comp}$) of 279 K is marked by the arrow. **c and d,** Magnetic structures of the high-temperature state (above $T_{comp}$) and the low-temperature state (below $T_{comp}$) of GdIG. For $T > T_{comp}$, the net magnetization of GdIG is dominated by the $d$-site Fe sublattice, while for $T < T_{comp}$, the net magnetization is dominated by the $c$-site Gd sublattice and the $a$-site Fe sublattice.

The $T_{comp}$ of the Pt/GdIG/MgO/PMN-PT sample is determined to be 279 K from its magnetization vs. temperature (*M-T*) curve (**Figure S1**b.), which is slightly lower than that of the bulk GdIG sample due to the Fe deficiency and lattice mismactch. [3,4] The magnetizations of the three sub-lattices in GdIG exhibit different temperature dependence, leading to the magnetic compensation transition. [5] At $T > T_{comp}$, the net magnetization of GdIG is dominated by the *d*-site Fe magnetization, which is along the external magnetic field direction but opposite to those of the Gd and the *a*-site Fe, as shown in **Figure S1**c. With decreasing temperature, the magnetizations of the *c*-site Gd and the *a*-site Fe increase more rapidly than that of the *d*-site Fe. Eventually, the sublattice magnetizations pointing in opposite directions cancel out at $T \approx T_{comp}$. When temperature further decreases to $T < T_{comp}$, the magnetizations of the *c*-site Gd and the *a*-site Fe become dominated and reverse their directions to align along the magnetic field direction, resulting in reduced Zeeman energy, as shown in **Figure S1**d.

Bauer, E. Saitoh, R. Gross, S. T. B. Goennenwein, M. Kläui, *Nat. Commun.* **2016**, 7, 10452.